# First results from the MIT Optical Rapid Imaging System (MORIS) on the IRTF: a stellar occultation by Pluto and a transit by exoplanet XO-2b

A. A. S. Gulbis[1,2], S. J. Bus[3], J. L. Elliot[2,4,5], J. T. Rayner[3], W. E. Stahlberger[3], F. E. Rojas[6], E. R. Adams[2], M. J. Person[2], R. Chung[3], A. T. Tokunaga[3], C. A. Zuluaga[2]

[1] Southern African Large Telescope and South African Astronomical Observatory, P.O. Box 9, Observatory, 7935 Cape Town, South Africa; amanda@saao.ac.za.

[2] Department of Earth, Atmospheric, and Planetary Sciences, Massachusetts Institute of Technology, 77 Massachusetts Ave. Cambridge, MA 02139; jle@mit.edu; era@mit.edu; mjperson@mit.edu; czuluaga@mit.edu.

[3] Institute for Astronomy, University of Hawaii, 640 N. A'ohoku Place #209, Hilo, HI 96720; sjb@ifa.hawaii.edu; rayner@ifa.hawaii.edu; stahlberw001@hawaii.rr.com; rchung@hale.ifa.hawaii.edu; tokunaga@ifa.hawaii.edu.

[4] Department of Physics, Massachusetts Institute of Technology, 77 Massachusetts Ave. Cambridge, MA 02139.

[5] Lowell Observatory, 1400 W. Mars Hill Rd., Flagstaff, AZ 86001.

[6] Department of Mechanical Engineering, Massachusetts Institute of Technology, 77 Massachusetts Ave. Cambridge, MA 02139; folkersr@mit.edu.

**ABSTRACT**. We present a high-speed, visible-wavelength imaging instrument: MORIS (the MIT Optical Rapid Imaging System). MORIS is mounted on the 3-m Infrared Telescope Facility (IRTF) on Mauna Kea, HI. Its primary component is an Andor iXon camera, a nearly 60 arcsec square field of view with high quantum efficiency, low read noise, low dark current, and full-frame readout rates ranging from as slow as desired to a maximum of between 3.5 Hz and 35 Hz (depending on the mode; read noise of 6e-/pixel and 49 e-/pixel with electron-multiplying gain=1, respectively). User-selectable binning and subframing can increase the cadence to a few hundred Hz. An electron-multiplying mode can be employed for photon counting, effectively reducing the read noise to sub-electron levels at the expense of dynamic range. Data cubes, or individual frames, can be triggered to several nanosecond accuracy using the Global Positioning System. MORIS is mounted on the side-facing exit window of SpeX (Rayner et al. 2003), allowing simultaneous near-infrared and visible observations. Here we describe the components, setup, and measured characteristics of MORIS. We also report results from the first science observations: the 24 June 2008 stellar occultation by Pluto and an extrasolar planetary transit by XO-2b. The Pluto occultation, of a 15.8 $R$ magnitude star, has signal-to-noise ratio of 35 per atmospheric scale height and a midtime error of 0.32 s. The XO-2b transit reaches photometric precision of 0.5 millimagnitudes in 2 minutes and has a midtime timing precision of 23 seconds.

# 1. INTRODUCTION

The observational frontier of astronomy has been expanded by the capability for increasingly shorter and more accurate time resolution. Technological advances in many areas have contributed to the efficacy of high-speed instrumentation: frame-transfer CCDs (charge-coupled devices) with effectively no deadtime, high quantum efficiency, and low read noise; increased computational data-streaming speeds and storage capacities; and portable, affordable access to the GPS (Global Positioning System). Sub-second observations have revealed new information about pulsars, cataclysmic variables, X-ray binaries, stellar pulsations (e.g. Phelan et al. 2008; Buckley et al. 2010), and bodies in the solar system through the observation of stellar occultations (e.g. Elliot 1979; Elliot & Olkin 1996).

In particular, sub-second resolution of occultations by bodies in the outer solar system probe these objects at the highest spatial resolution that can be achieved by any Earth-based method (i.e. a few kilometers at 30 AU: Gulbis et al. 2006; Person et al. 2006). Recent observations of stellar occultations by Pluto indicate that its atmosphere has halted its expansion (Young et al. 2006; Elliot et al. 2007) and reveal dynamical features such as high-altitude waves (McCarthy et al. 2008; Person et al. 2008). Continued monitoring via occultation observations is necessary to assess atmospheric changes as Pluto moves away from perihelion and before the arrival of NASA's *New Horizons* spacecraft in 2015.

A similar technique has been used to discover and characterize planets outside of the solar system: observing the fractional change in light as an exoplanet transits in front of its parent star. While only a subset of exoplanets are geometrically positioned for these events to be viewed from Earth, there are now over 100 known transiting exoplanets (see http://exoplanet.eu/catalog.php for the most recent count). High-quality photometry and accurate timing of exoplanet transits provides information on the planet's mass, the ratio between the planetary and stellar radius, the semimajor axis of the planet, stellar limb darkening parameters, and can reveal transit timing variations that may be due to additional planets or satellites.

Here we present a new optical, high-time-resolution instrument called MORIS (the MIT Optical Rapid Imaging System) and its first science observations of a stellar occultation and an exoplanet transit. MORIS is located on NASA's 3-m Infrared Telescope Facility (IRTF) on Mauna Kea, Hawai'i. It is mounted on the side-facing exit window of SpeX, a 0.8-5.4 μm spectrograph and imager (Rayner et al. 2003). This mounting allows for simultaneous high-speed observations at visible wavelengths (using MORIS) and in the near-IR (using SpeX).

In Section 2, the instrument and its components are described in detail. Section 3 contains results from a single-chord observation of a stellar occultation by Pluto on 2008 June 24 and a transit by exoplanet XO-2b on 2008 December 06. A summary and future work are presented in Section 4.

## 2. INSTRUMENT SPECIFICATIONS

MORIS is based on the Portable Occultation, Eclipse and Transit Systems (POETS: Souza et al. 2006; Gulbis et al. 2008). POETS were developed to allow high-quality, high-speed, GPS-timed, optical observations in an easy-to-transport format. MORIS has a similar design to POETS; however, it is optimized to interface with SpeX and is permanently based at the IRTF. The primary instrument components are described in detail below. A schematic of the instrument setup is shown in Figure 1, and a summary of the instrument characteristics is provided in Table 1.

## 2.1. Camera

The backbone of the system is an Andor iXon$^{EM+}$ DU-897 camera. This is a frame-transfer, high-quantum-efficiency, low-read-noise camera. A unique aspect of this camera is the selection of either conventional readout or electron multiplying (EM) via an extended serial register prior to readout. In EM mode, transferred electrons undergo impact ionization, strengthening the observed signal without increasing read noise. This process effectively reduces read noise to sub-electron levels. The output of either the conventional or EM register is routed to a preamplifier stage at which one of three gain settings is selected (1×, 2.4×, or 5×). The signal is then fed to one of two Analog to Digital Converters (ADC): a 16-bit ADC with readout rate 1 MHz or a 14-bit ADC with readout rates 3, 5, or 10 MHz. The 1 and 3 MHz rates are available in conventional mode and all rates function in EM mode. Each mode, preamp setting and readout rate has different performance characteristics. See Table 2 for read noise and gain measurements.

A measurement of the linearity of the detector is shown in Figure 2. We find that for the three tested modes (two conventional and one EM), the linearity conforms to the manufacturer's specification of being better than 1% over the full range of counts. Depending on the mode, saturation is reached at either the ADC bit limit or the full-well capacity.

The benefit of EM can be significant. However, three important factors must be considered when using this mode: dynamic range, excess noise, and clock-induced charge. First, dynamic range in conventional modes ranges between approximately 10,000 and 30,000, depending on the ADC and preamplifier gain setting. For EM modes, the dynamic range increases at low EM gain as read noise is effectively reduced (EM=1 to EM≈10). Once the register can no longer accommodate the amplified signal, the dynamic range levels out (EM≈10 to EM≈40). When the effective read noise falls below single photon levels, the dynamic range drops with increasing EM gain (EM≈40 to EM≈1000). Quantitatively, the EM mode dynamic range starts between approximately 2000 and 9000 (depending on the ADC and preamplifier gain settings), levels off at approximately15 bits for 1 MHz, 14 bits for 5 MHz, and 13 bits for 10 MHz, and then drops as the inverse of the gain down to a few hundred by EM=1000. Second, the EM mode generates an excess noise factor of up to $\sqrt{2}$ due to the stochastic nature of the EM amplification process (e.g. Basden et al. 2003; Robbins & Hadwen 2003). Third, clock-induced-charge (electrons generated during readout) is amplified by the EM register and can become a significant factor at high EM gain settings (e.g. Daigle et al. 2009).

The camera is cooled thermoelectrically. Under normal operations, MORIS is air-cooled to –70º C. Water cooling could be employed to reach temperatures on the order of –100º C. We have not found water cooling to be necessary for standard usage.

## 2.2. Fore optics Box

SpeX has two $CaF_2$ dichroic beam splitters, which cut the beam at either 0.8 μm or 0.95 μm – wavelengths shorter than these can be directed into the MORIS optical path. The beam exiting from SpeX is f/38. To speed up the beam and allow stronger signal over shorter time, we chose a focal reduction of 3:1. There is a constraint on the distance that the instrument can extend from the side of SpeX to allow adequate clearance for instrument swapping at the IRTF Cassegrain focus. Thus, a folding mirror is required to

*accepted in Publications of the Astronomical Society of the Pacific*     6

direct the beam downward. The path was also designed to minimize the number of optics in order to prevent loss of light. A schematic of the resulting optical path is shown in Figure 3. This three-lens, one-mirror design was created using the ZEMAX optical design program from Focus Software, Inc.

The three lenses are custom made, the first and third from Schott SF5 and the second from Schott BK7. All lenses are anti-reflectance coated with a single layer of $MgF_2$ at 590 nm to a level of < 3.5% reflectance over visible wavelengths. The mirror is coated with protected silver, allowing > 90% reflectivity over visible wavelengths. The fused-silica camera window has a standard Andor anti-reflectance coating, with reflectance of roughly 4% over visible wavelengths.

To derive the response of the system, we consider the CCD quantum efficiency and the throughputs of the components in the fore optics box as described above. We combine these values with the throughputs of the primary mirror, secondary mirror, SpeX dichroic, and two SpeX windows with single-layer $CaF_2$ coating (efficiencies given in Table 4 of Rayner et al. 1993). The total MORIS detector response, for each dichroic, is shown in Figure 4a.

There is a custom-made filter wheel that holds ten, 1-inch diameter filters. The wheel is rotated by an Animatics SmartMotor SM2330D, and a positive mechanical detent locking system is employed to ensure positioning repeatability to better than 15 μm. The current suite of 5-mm thick filters were made by Asahi Spectra: SDSS $g'$, $r'$, $i'$, and $z'$ (following the Sloan Digital Sky Survey, Fukugita et al. 1996), Johnson *V*, *VR* (following Jewitt et al. 1996), and OG590 (a longpass-red filter comprised of 3 mm Schott OG590 and 2 mm BK7). The detector response including the transmission curves of the filters is plotted in Figures 4b and 4c. ZEMAX simulations show that there should be a negligible change in focus between the different filters or an open position.

We designed a light-tight box to interface with SpeX, house the optics and filter wheel, and allow mounting and focusing of the camera. The anodized components of the fore optics box were machined at MIT to a positional accuracy of better than 0.005 inches. Figure 5 is a photograph of this box mounted onto the side of SpeX.

**2.3. Timing**

To ensure accurate timing, we use a Spectrum Instruments, Inc. Intelligent Reference/TM-4 GPS. A small weather-proof antenna has been attached to the top of the telescope with an RG-58 coaxial cable trailing down to the GPS, which is located in an IRTF cool rack. Four satellites are required to derive accurate latitude, longitude, altitude, and time. After the position is established, only one satellite is required to derive a precise time. During all MORIS engineering and observing runs to date, adequate satellite signals have been obtained. This system is accurate to well under 1 μsec (the manufacturer specification is ±25 nsec from UTC, with root-mean-square stability of 12.5 nsec).

In addition to a standard one pulse-per-second output, the GPS has an output for a programmable pulse. This can be set to one pulse of fixed width or as a series of perpetual pulses at a specified interval and having specified width. The camera can thus receive an input pulse from the GPS that will begin a series of camera-controlled exposures, or the readout of each exposure can be triggered by a separate input pulse. The latter of these two modes is typically employed, via an RG-174 coaxial cable of

*accepted in Publications of the Astronomical Society of the Pacific*     8

known length and thus known timing delay, in order to ensure the most accurate timing and cadence for large data cubes.

### 2.4. Control computer

MORIS is operated by a custom Shuttle XPC SD30G2 small form factor computer. It has a dual-core 3.2GHz Intel Pentium processor, with 2GB RAM. There are two 150GB, 10,000 RPM Western Digital Raptor hard drives. These parameters are required to allow high image acquisition rates (Souza at al. 2006). A roughly three-quarter-length PCI card serves to interface with the camera, and the GPS is connected via a serial port.

The camera is controlled using Andor's Solis (Solutions for Imaging and Spectroscopy) software. The GPS is controlled using Spectrum Instrument's control/display software. Each of these programs is Windows-based, so the computer runs on a 32-bit Microsoft Windows XP Professional platform. All software is installed on both hard drives such that there is a backup in case of a primary disk failure.

## 3. SCIENCE OBSERVATIONS

### 3.1. Stellar occultation of P571 by Pluto

MORIS was mounted on the IRTF in May 2008. Our first observing run was on the night of 23/24 June to view the predicted stellar occultation of P571 (USNO-B 0729-0691269; *R* magnitude 15.8) by Pluto (McDonald & Elliot 2000). The refined occultation prediction was based on nearly 50 astrometric observations of the star and more than 300 astrometric observations of Pluto: the measured star position showed an offset of nearly +0.4 arcsec in RA and –0.2 arcsec in declination from the catalog position: R.A. =

17:58:22.3951 and Dec. = –17:02:49.347 at the epoch of the event (http://occult.mit.edu/research/occultations/Pluto/P571-preds/index.html). The predicted shadow path had a center line 120 ± 443 km North of Mauna Kea and the predicted midtime for the occultation from the IRTF was 24 June 2008 10:35:53 ± 00:00:10 UT. The IRTF was the only site from which we arranged to observe the occultation.

The observations were carried out using the IRTF's remote observing option, which was economical given that we were awarded only four hours of telescope time to observe the event. The occultation data cube spanned 40 minutes, centered on the predicted midtime. Each frame was triggered from the GPS at a cadence of 4 Hz. The camera settings were 1 MHz conventional mode, with 2.4× preamplifier gain. We used the SpeX 0.95-μm dichroic with no filter. Binning was set to 2×2, providing an effective pixel size of 0.23 arcsec. There were thin clouds throughout the observations, the airmass ranged from 1.25 to 1.45, and the seeing was approximately 0.7 to 1.0 arcsec. Simultaneous observations were started using SpeX with Guidedog (Rayner et al. 2003); however, Guidedog crashed ten minutes before the predicted midtime and could not be restarted in time to observe the occultation.

Careful data reduction was required for this event because of the variable clouds. In addition, the MORIS field of view was vignetted at the corners and contained significant scattered light (Figure 6; see Section 4 for additional discussion). We performed aperture photometry on each frame to extract the combined signal from Pluto, Charon, and P571. This procedure was repeated for two comparison stars in the frame that were slightly brighter than P571. Background regions were carefully selected around each aperture in order to accurately subtract out background light. To calibrate the light curve, we used data taken before and after the occultation when Pluto and the star were well separated. We find the background fraction, (Pluto+Charon)/(Pluto+Charon+P571), to be 0.826 ± 0.001. The highest signal-to-noise ratio (SNR) for the unocculted Pluto, Charon, and P571 baseline was obtained using raw data, with a circular aperture of 12 superpixels diameter, and dividing by the mean of the signal from the two comparison stars. The calibrated lightcurve shown in Figure 7 has SNR of 35 per atmospheric scale height of 60 km on Pluto.

A geometric solution was not possible since we have only a single chord for this occultation. In addition, extensive attempts to derive object positions using multiple point-spread-function fitting did not return accurate results. A significantly larger number of astrometric frames, extending further before and after the event, would be required for that analysis.

Following Elliot & Young (1992), we fit the lightcurve by assuming a power-law thermal structure. We assumed the best-fit atmospheric parameters from Elliot et al. (2007) of half-light radius $r_H$ = 1276.1 km, equivalent isothermal gravitational to thermal energy ratio at half light-radius $\lambda_{iso}$ = 18.3, and thermal gradient parameter $b$ = -2.2. A one-limb model was used, and the surface radius was set to a low enough value such that it would not effect the light curve. Only data points of flux > 0.4 were used, since Pluto's lower atmosphere is known to deviate from isothermal (e.g. Elliot et al. 2007; Young et al. 2008, and references therein; Lellouch et al. 2009). Best-fit values were found for closest approach distance (99.9 ± 99.9 km), midtime (1143.80 ± 0.32 seconds from the datacube start time of 10:18:00.00 UT), full-scale unoccultated signal (1.02 ± 0.49), and slope of the full-scale signal (–5.49 × $10^{-6}$ ± 3.34 × $10^{-4}$). The best-fit model is plotted along with the data in Figure 7. Although some structure is apparent at the base of the light curve, the large central flash predicted by the model is not evidenced. A thermal gradient and/or extinction in the lower atmosphere could cause this feature to be suppressed.

The error on the predicted closest approach was 99.9 km, or 0.005 arcsec, which compares favorably with errors on recently observed stellar occultations by Pluto (e.g. errors on predicted closest approaches of 0.010 and 0.015 arcsec in 2006 and 2007 respectively; Elliot et al. 2007; Person et al. 2008). The best-fit midtime is 10:37:03.80 ± 0.32 UTC. The error on the predicted midtime for this station was thus 00:01:11, which also compares favorably to previous predictions and observations (e.g. errors on predicted midtimes of 00:02:33 and 00:02:24 in 2006 and 2007 respectively; Elliot et al. 2007; Person et al. 2008).

### 3.2. Extrasolar planetary transit by XO-2b

On 06 December 2008, MORIS was used to observe a transit by the extrasolar planet XO-2b. This exoplanet was announced by Burke et al. (2007) with eleven light curves, and an additional six, high-quality light curves were provided by Fernandez et al. (2009). XO-2b was chosen to see how well MORIS would perform on a bright ($I = 10.5$ magnitude) exoplanet target with a binary companion. In this case, the companion is a nearly-identical star separated by 30 arcsec: an ideal situation for differential aperture photometry with MORIS.

The observations were carried out using the IRTF's remote observing option, with the SpeX 0.95-μm dichroic and a long-pass-red, visitor, Thor Labs filter (lower cutoff at 700 nm). MORIS was internally triggered to take two-second exposures. The camera settings were full frame, 1 MHz conventional mode, with 2.4× preamplifier gain. Throughout the observations, the airmass ranged from 2.2 to 1.2 and the seeing was approximately 1 arcsec. The best light curve, shown in Figure 8, was produced with a 30-pixel radius aperture around the star, with the sky background estimated from an annulus with inner radius of 60 pixels and a width of 10 pixels. A slight mismatch in the level of baseline before and after transit was removed with a trend against time. The instrument was refocused twice when the seeing began to worsen, and there are two features just before ingress and near the midtime associated with those refocusing events.

The transit light curve was fit using the Mandel & Agol (2002) algorithm as implemented by the white-noise model described in Carter & Winn (2009). We assumed that XO-2b has zero obliquity, oblateness, and orbital eccentricity, and we employed a quadratic limb-darkening law assuming $T = 5340$ K, $\log g = 4.48$, $[M/H] = 0.5$, and $V_{micro} = 2$ km/s. Since the transit was observed with a longpass filter (effective wavelength 700-900 nm), we used the initial values of $u_1 = 0.3670$ and $u_2 = 0.2850$ for the Sloan $i'$ filter from Claret (2004) and fit for the linear term, $u_1$, while leaving $u_2$ fixed. The best model parameters were found using a Monte Carlo Markov chains (MCMC) method, with the best least-squares-fit values used as initial parameters. Three independent chains of $10^6$ links (minus the first 50,000 points in each) were combined to derive the final parameters, which are the median and 68.3% credible interval values (equivalent to the standard deviation if the distribution is Gaussian). For more details on the modeling, see Adams (2010). Following Fernandez et al. (2009), the orbital period was fixed to $P = 2.615864$ days, and we assumed that $M_* = 0.971 \pm 0.034\ M_{sun}$, $R_* = 0.976 \pm 0.024\ R_{sun}$, and $M_{planet} = 0.565 \pm 0.054\ M_{Jupiter}$.

*accepted in Publications of the Astronomical Society of the Pacific*    13

By fitting the transit on 2008 December 06 jointly with another half-transit observed from MORIS on 2008 December 19, we derived a planetary radius of $R_{planet}$ = 0.955 ± 0.024 $R_{Jupiter}$. This result is slightly smaller than, but consistent with, the value of $R_{planet}$ = 0.996 $^{+0.031}_{-0.018}$ $R_{Jupiter}$ reported by Fernandez et al. (2009). Additional fit parameters are described in Adams (2010). Correlated noise is present in the light curve, most notably as a sharp increase just after midtransit that was associated with a refocusing event and also coincided with a slight positional change. Consequently, the formal fit errors were inflated by a factor of 3.1 based on an examination using the time-averaged residual method (Pont et al. 2006; Adams et al. 2010). The measured midtime, 2454806.94750 ± 0.00027 $BJD_{TDB}$, agrees with the predicted midtime of Fernandez et al. (2009) and no signs of transit timing variation are seen in this light curve. We reached a photometric precision of 0.5 mmag in 2 minutes and a midtime timing precision of 23 seconds.

## 4. SUMMARY AND FUTURE WORK

This paper introduces a new, high-speed, accurately-timed, optical imaging camera system on NASA's 3-m IRTF. The IRTF's remote observing capabilities make this instrument particularly ideal for targeted, time-sensitive observations, two examples of which are presented here. The observations of a stellar occultation by Pluto and an exoplanet transit by XO-2b demonstrate that the system is capable of taking well-timed, high-quality, photometric data. First, an occultation lightcurve of decent SNR (35 per scale height) was obtained on a fairly faint star (15.8 *R* magnitude), and the midtime has an accuracy of 0.32 s. The results indicate that the lightcurve may contain a repressed central flash and that the prediction was extremely accurate. If chords from other stations would have been obtained, these data would be of sufficient quality characterize Pluto's atmosphere. Given Hawaii's unique geographical location, we anticipate that MORIS will play a key role in future occultation observations by Pluto and other large Kuiper Belt objects (e.g. Elliot et al. 2010). Second, the transit data reach the millimag photometric accuracy required for detection and have a timing precision of tens of seconds. Thus, MORIS observations can achieve the level of accuracy needed to reduce the ambiguity between correlated parameters such as orbital inclination, radius ratio, and stellar limb darkening and can possibly reveal variations caused by additional bodies in an extrasolar planetary system.

A key strength of MORIS is the ability to take simultaneous NIR observations with SpeX. Although the results presented here did not make use of that feature, we will explore it in future work. Such multi-wavelength observations will be important to occultations because they allow distinction between atmospheric extinction and differential refraction (Elliot et al. 2003). For transit observations, simultaneous multi-wavelength observations can help constrain modeling parameter space, provide independent confirmation of light curve timing, and potentially reveal wavelength-dependent features (e.g. Colon et al. 2010). Similarly, the electron multiplying capability of the system could provide enhanced results and will be studied in future work.

In the presented results, we do not address flatfield calibrations. MORIS exhibited a significant (factor of two) radial variation in brightness, the exact structure of which varied as a function of pointing and ambient illumination. The IRTF is optimized for infrared observations rather than visible, and we have confirmed that the background brightness effect was the result of scattered light. We designed a new optical path and foreoptics box, which include a stop and multiple baffles that effectively eliminate all light that is not from the secondary mirror. We will present details of the modified instrument in a subsequent paper.

Funding for this work was provided by NASA PA/PME grant NNX07AK95G. A.A.S.G., E.R.A., and M.J.P. were visiting astronomers at the Infrared Telescope Facility, which is operated by the University of Hawaii under Cooperative Agreement no. NNX-08AE38A with the National Aeronautics and Space Administration, Science Mission Directorate, Planetary Astronomy Program.

**REFERENCES**

Adams, E. R. 2010, Transit Timing with Fast Cameras on Large Telescopes, Ph.D. thesis, Massachusetts Institute of Technology

Adams, E. R., Lopez-Morales, M., Elliot, J. L., Seager, S., & Osip, D. J. 2010, ApJ, 714, 13

Basden, A. G., Haniff, C. A., & Mackay, C. D. 2003, MNRAS, 345, 985

Buckley, D. A. H., et al. 2010. in Ground-based and Airborne Instrumentation for Astronomy III (Proceedings of SPIE), Time Resolved Astronomy with the SALT, ed. I. S. M. S. K. R. H. Takami (San Diego, California)

Burke, C. J., et al. 2007, ApJ, 671, 2115

Carter, J. A., & Winn, J. N. 2009, ApJ, 704, 51

Claret, A. 2004, Astronomy and Astophysics, 428, 1001

Colon, K. D., Ford, E. B., Lee, B., Mahadevan, S., & Blake, C. H. 2010, MNRAS, in press

Daigle, O., Carignan, C., Gach, J.-L., Guillaume, C., Lessaed, S., Fortin, C.-A., & Blais-Ouellette, S. 2009, PASP, 121, 866

Elliot, J. L. 1979, ARA&A, 17, 445

Elliot, J. L., et al. 2003, Nature, 424, 165

Elliot, J. L., & Olkin, C. B. 1996, in Annual Review of Earth and Planetary Sciences, eds. G. W. Wetherill, A. L. Albee, & K. C. Burke (Palo Alto: Annual Reviews Inc.), 89

Elliot, J. L., et al. 2007, AJ, 134, 1

---. 2010, Nature, 465, 897

Elliot, J. L., & Young, L. A. 1992, AJ, 103, 991

Fernandez, J. M., Holman, M. J., Winn, J. N., Torres, G., Shporer, A., Mazeh, T., Esquerdo, G. A., & Everett, M. E. 2009, AJ, 137, 4911

Fukugita, M., Ichikawa, T., Gunn, J. E., Doi, M., Shimasaku, K., & Schneider, D. P. 1996, AJ, 111, 1748

Gulbis, A. A. S., et al. 2006, Nature, 439, 48

Gulbis, A. A. S., Elliot, J. L., Person, M. J., Babcock, B. A., Pasachoff, J. M., Souza, S. P., & Zuluaga, C. A. 2008. in The Universe at Sub-second Timescale, High Time Resolution Astrophysics, Recent Stellar Occultation Observations Using High-Speed, Portable Camera Systems, eds. D. Phelan, O. Ryan, & A. Shearer (Edinburgh, Scotland: American Institute of Physics), 91

Jewitt, D., Luu, J., & Chen, J. 1996, AJ, 112, 1225

Lellouch, E., Sicardy, B., de Bergh, C., Käufl, H.-U., Kassi, S., & Campargue, A. 2009, Astronomy and Astophysics, 495, L17

Mandel, K., & Agol, E. 2002, ApJ, 580, L171

McCarthy, D., Kulesa, C., Hubbard, W., Kern, S. D., Person, M. J., Elliot, J. L., & Gulbis, A. A. S. 2008, AJ, 136, 1519

McDonald, S. W., & Elliot, J. L. 2000, AJ, 120, 1599

Person, M. J., Elliot, J. L., Gulbis, A. A. S., Pasachoff, J. M., Babcock, B. A., Souza, S. P., & Gangestad, J. W. 2006, AJ, 132, 1575

Person, M. J., et al. 2008, AJ, 136, 1510

Phelan, D., Ryan, O., & Shearer, A. 2008, in Astrophysics and Space Science Library (Dordrecht: Springer Netherlands), 350

Pont, F., Zucker, S., & Queloz, D. 2006, MNRAS, 373, 231

Rayner, J., et al. 1993, Proc. SPIE, 1946, 490

Rayner, J. T., Toomey, D. W., Onaka, P. M., Denault, A. J., Stahlberger, W. E., Vacca, W. D., Cushing, M. C., & Wang, S. 2003, PASP, 115, 362

Robbins, M. S., & Hadwen, B. J. 2003, IEEE Transactions on Electron Devices, 50, 1227

Souza, S. P., Babcock, B. A., Pasachoff, J. M., Gulbis, A. A. S., Elliot, J. L., Person, M. J., & Gangestad, J. W. 2006, PASP, 118, 1550

Young, E. F., et al. 2008, AJ, 136, 1757

Young, L., Buie, M., French, R., Olkin, C., Regester, J., Ruhland, C., & Young, E. 2006, BAAS, 38, 542

TABLE 1
SUMMARY OF INSTRUMENT CHARACTERISTICS

| | |
|---|---|
| Final beam speed | f/12.7 |
| CCD | E2V CCD97; 512 × 512, 16μm$^2$ pixels |
| Full well capacity [a] | 158,268 e- |
| Linearity [a] | < 1% |
| Plate scale [b] | 0.1139 arcsec/pixel |
| Field of view [b] | 58.3 arcsec × 58.3 arcsec |
| Readout rate | 3.5 frames/sec (full frame, 1MHz amplifier); 35 frames/sec (full frame, 10 MHz amplifier); up to hundreds of frames/sec with binning and/or subframing |
| Dead time | 1.7 msec (for 512 rows at the default vertical shift speed) |
| Dark current [c] | < 0.001 e-/pix/sec |
| GPS accuracy | < 1 μsec |
| GPS antenna cable delay | 77 nsec [d] |
| GPS trigger cable delay | 30.8 nsec [d] |
| Current filters | SDSS $g'$, $r'$, $i'$, and $z'$; Johnson $V$; $VR$; OG590 (long-pass red) |

[a] The saturation signal per pixel and linearity up to saturation (as a percentage variation from a straight–line fit) provided by the manufacturer. Linearity measurements for three different camera modes are provided in Figure 2.

[b] The plate scale was measured from astrometric fits using stars from the UCAC2 catalog to images of open cluster Berkeley 81.

[c] Dark current is negligible in exposures up to 240 sec at –70º C. This value represents the manufacturer's specification.

[d] Based on a manufacturer-specified time delay of 1.54 nsec/ft, for 50 feet of RG-58 and 20 feet of RG-174 coaxial cable.

TABLE 2

MEASURED READ NOISE AND GAIN

| | Readout amplifier and preamplifier gain | | | | | | | | | | | |
|---|---|---|---|---|---|---|---|---|---|---|---|---|
| **Conventional Mode** | 1MHz[a] | | | 3MHz | | | | | | | | |
| | *1×* | *2.4×* | *5×* | *1×* | *2.4×* | *5×* | | | | | | |
| Read noise (e-)[b] ...... | 8.58 | 6.35 | 5.83 | 13.56 | 10.04 | 9.49 | | | | | | |
| Gain (e-/ADU)[b] ...... | 3.70 | 1.47 | 0.66 | 9.78 | 3.89 | 1.78 | | | | | | |
| **EM Mode (EM=3)** | 1 MHz[a] | | | 3 MHz | | | 5MHz | | | 10 MHz | | |
| | *1×* | *2.4×* | *5×* | *1×* | *2.4×* | *5×* | *1×* | *2.4×* | *5×* | *1×* | *2.4×* | *5×* |
| Read noise (e-)[b] ...... | 8.98 | 5.49 | 4.41 | 12.37 | 7.49 | 6.30 | 11.68 | 7.33 | 2.18 | 12.33 | 7.70 | 5.88 |
| Gain (e-/ADU)[b] ...... | 4.55 | 1.83 | 0.82 | 10.99 | 4.47 | 2.05 | 7.33 | 3.26 | 1.48 | 8.25 | 1.76 | 0.74 |
| **EM Mode (EM=40)** | 1 MHz[a] | | | 3 MHz | | | 5MHz | | | 10 MHz | | |
| | *1×* | *2.4×* | *5×* | *1×* | *2.4×* | *5×* | *1×* | *2.4×* | *5×* | *1×* | *2.4×* | *5×* |
| Read noise (e-)[b] ...... | 0.56 | 0.36 | 0.30 | 0.78 | 0.47 | 0.40 | 1.05 | 0.66 | 0.19 | 1.10 | 0.70 | 0.48 |
| Gain (e-/ADU)[b] ...... | 0.29 | 0.12 | 0.06 | 0.69 | 0.28 | 0.13 | 0.70 | 0.29 | 0.13 | 0.74 | 0.31 | 0.14 |
| **EM Mode (EM=100)**[c] | 5MHz | | | 10MHz | | | | | | | | |
| | *1×* | *2.4×* | *5×* | *1×* | *2.4×* | *5×* | | | | | | |
| Read noise (e-)[b] ...... | 0.42 | 0.26 | 0.08 | 0.45 | 0.28 | 0.18 | | | | | | |
| Gain (e-/ADU)[b] ...... | 0.28 | 0.12 | 0.05 | 0.30 | 0.12 | 0.05 | | | | | | |

[a] Amplifier for these modes is 16 bit. All other modes are 14 bit.
[b] Effective read noise and gain values are the mean of multiple measurements with the camera air-cooled to –70ºC.
[c] Values at a higher EM setting are given only for 5MHz and 10 MHz amplifiers to demonstrate subelectron read noise for all gain settings.

**FIGURE CAPTIONS**

Figure 1. Schematic drawing of the MORIS setup. The "cool rack" refers to one of four thermally-isolated cabinets around the perimeter of the IRTF Multiple Instrument Mount. The IRTF ethernet, and switch in the Telescope Control System (TCS) room, are employed to route the control of the instrument to one of the displays in the observer's area or remotely via VNC.

Figure 2. MORIS linearity measurements of an evenly illuminated field. The average, bias-subtracted counts per pixel are plotted versus time, along with best-fit lines. Three modes were tested: 1 MHz conventional 2.4× (16-bit; 1.47 e-/ADU), 3 MHz conventional 2.4× (14-bit; 3.89 e-/ADU), and 5 MHz EM off 2.4× (14-bit; 23.05 e-/ADU). Horizontal dashed lines represent saturation limits. The conventional modes reach saturation due to the data transfer limitation of the ADC, while the EM mode reaches saturation slightly past the manufacturer's specified full-well capacity. All data lie within 1% of the best-fit lines.

Figure 3. ZEMAX diagram of the MORIS optical path, from the exit window of SpeX to the Andor Ixon camera window. The 3:1 reducing optics consist of three lenses and one folding mirror.

Figure 4. The MORIS detector response as a function of wavelength. The response is shown as a percentage of light entering the telescope and thus includes throughput from all optics as well as CCD quantum efficiency. (a) Response with no filter. The two

curves correspond to the SpeX 0.8-μm and 0.95-μm dichroic beam splitters, one of which must be selected to direct light out of the side-facing exit window and into the MORIS fore optics box. (b) Response for the Sloan filters, with the 0.95-μm dichroic. Sloan $g'$, $r'$, $i'$, and $z'$ are represented by dotted, gray, black, and dashed lines respectively. (c) Response for the remaining filters, with the 0.95-μm dichroic. Johnson *V*, *VR*, and longpass red (OG590) are represented by thick black, dot-dashed, and thick gray lines respectively.

Figure 5. MORIS mounted on the IRTF. The fore optics box is anodized black, and the gray Andor camera extending from the bottom of the box is protected by a black frame. The system is mounted on the side-facing exit window of the SpeX cryostat (blue box). The red handle (upper left) allows manual control of a shutter in front of the SpeX exit window.

Figure 6. Raw MORIS image of the P571 field. This is a 5-second exposure, binned 2x2, taken an hour before the occultation. Pluto, its moon Charon (resolved to the lower left of Pluto), the occultation star (P571), and the comparison stars used to generate the light curve are all labeled. The image has been scaled to display ADU 0 to 20,000 in order to highlight fainter features. The two comparison star peak counts are on the order of 50,000 ADU/pixel and Pluto's peak counts are approximately 40,000 ADU/pixel. The camera setting for this image was 1MHz conventional 2.4×, corresponding to gain 1.47 e-/ADU. A slight vignetting by one of the optics holders is apparent at the corners of the frame. In addition, there is a gradient in the background light that decreases radially by a factor of approximately 1.5 from the center to the edges.

Figure 7. Light curve from the occultation of P571 by Pluto. The normalized flux of the star is plotted versus time. (*top*) Data points are shown at full time resolution of 0.25 seconds. The thick gray line represents the model fit to flux greater than 0.4. The thin gray line shows the continuation of the model to lower flux (unfitted data): it is apparent that the lower atmosphere differs significantly from the model. (*bottom*) Data points are binned by 8 for a resolution of 2 seconds. The binned data are displayed to more clearly distinguish the features at the bottom of the light curve, which are possibly diagnostic of a repressed, asymmetric central flash.

Figure 8. Light curve from a transit of exoplanet XO-2b. (*top*) The normalized flux of the star is plotted versus time. Black dots represent data binned by ten seconds and the gray line is the model fit. (*bottom*) The residuals between the data and the model are plotted versus time, with a gray line at 0.

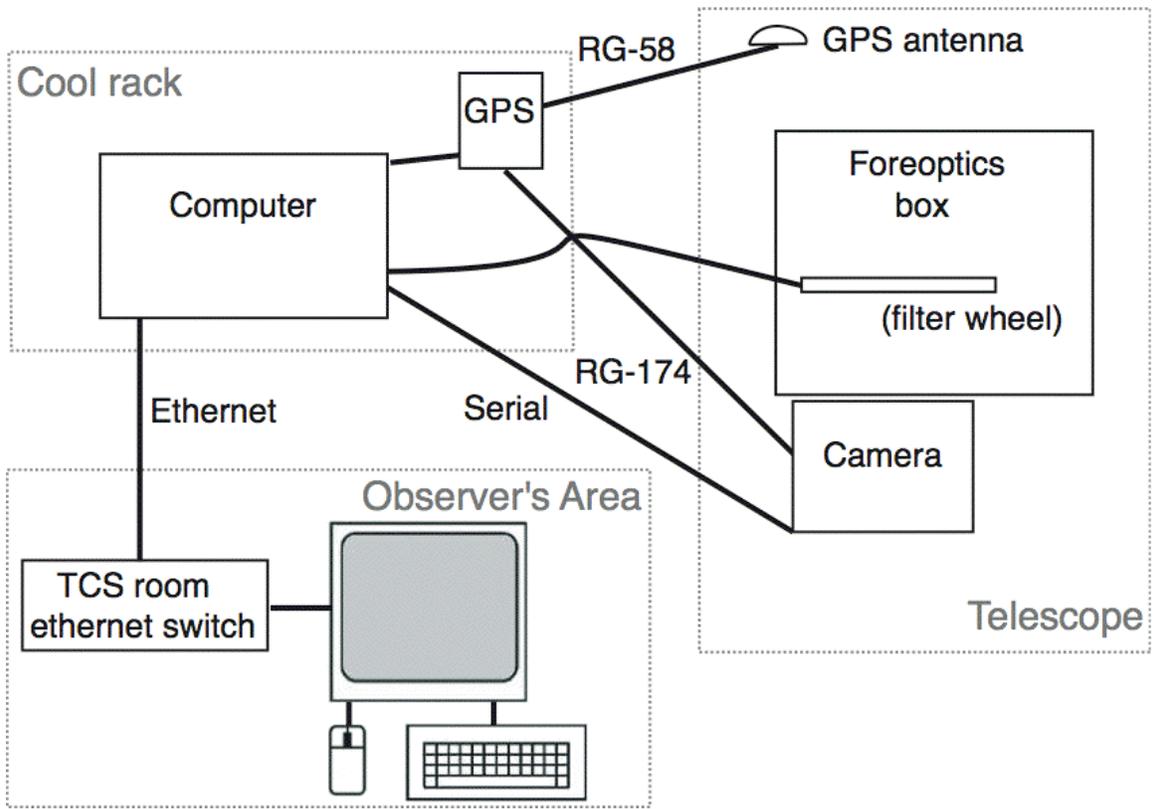

Figure 1. Gulbis *et al*.

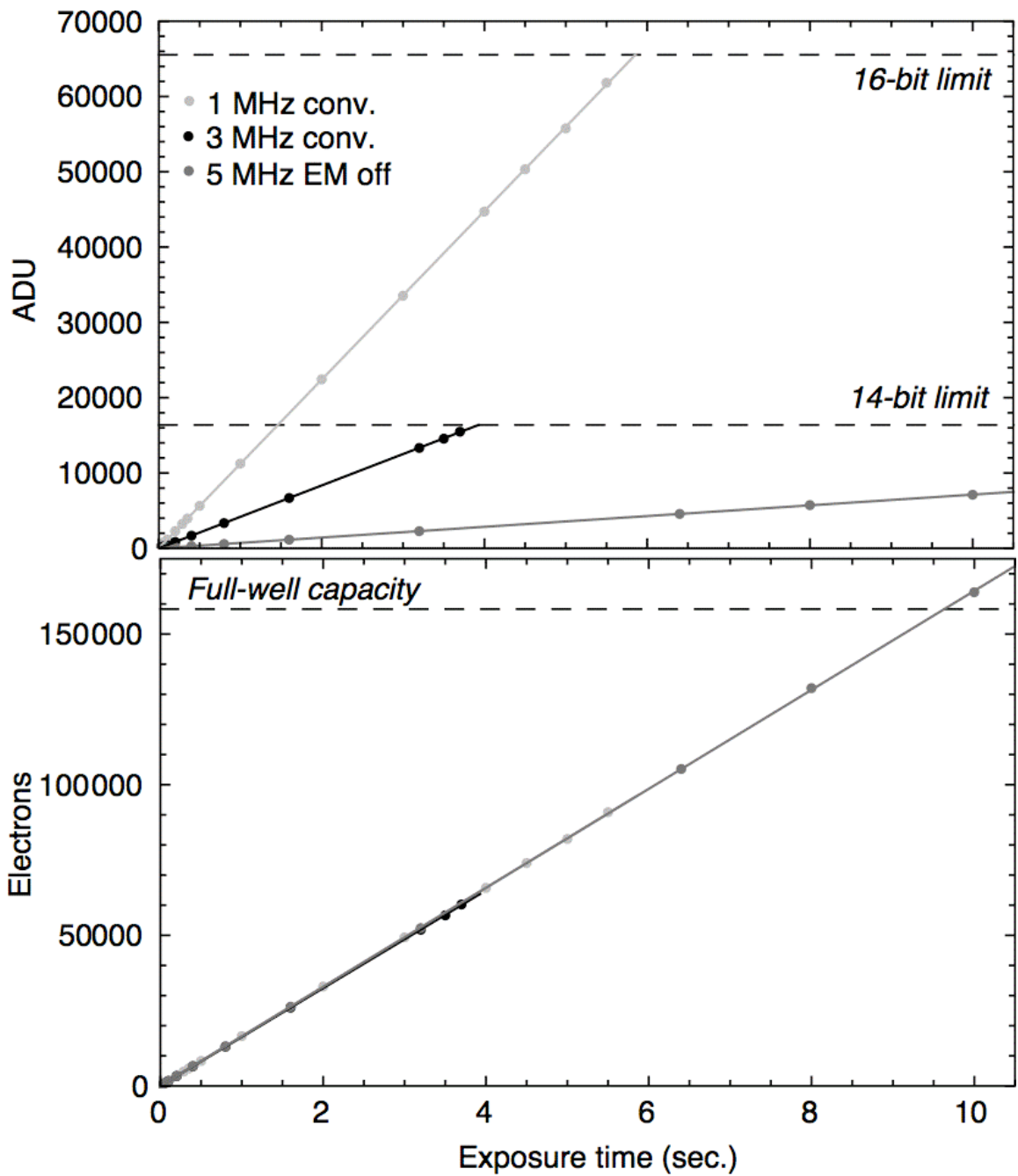

Figure 2. Gulbis *et al*.

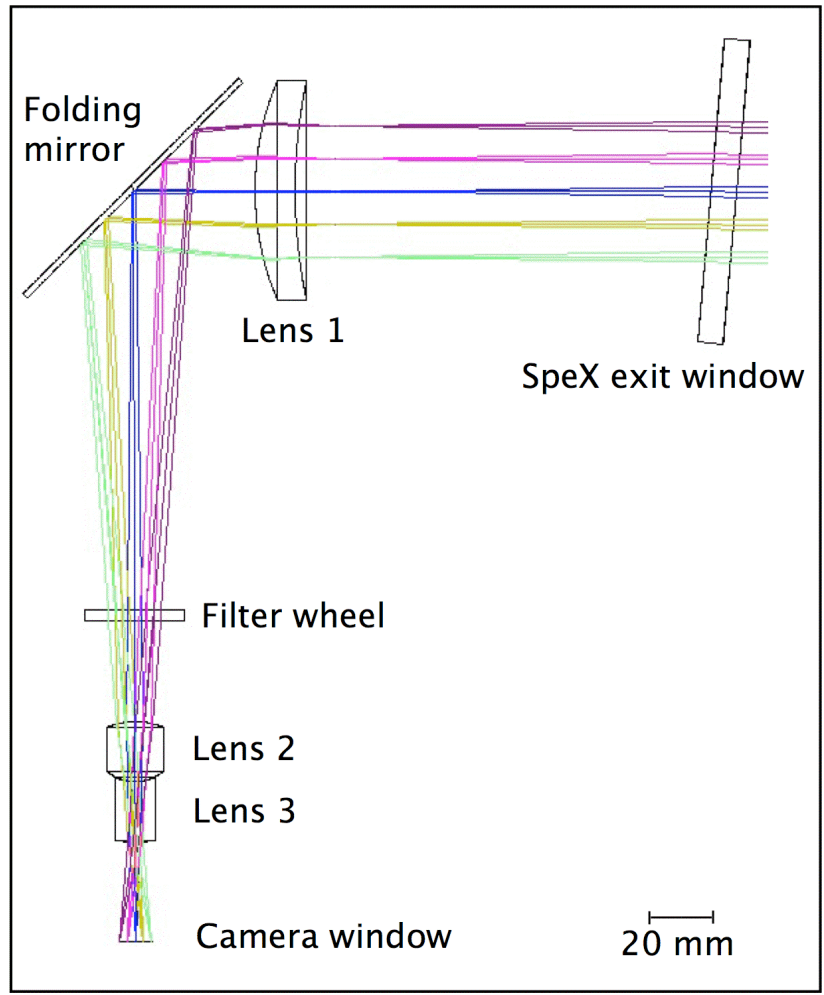

Figure 3. Gulbis *et al*.

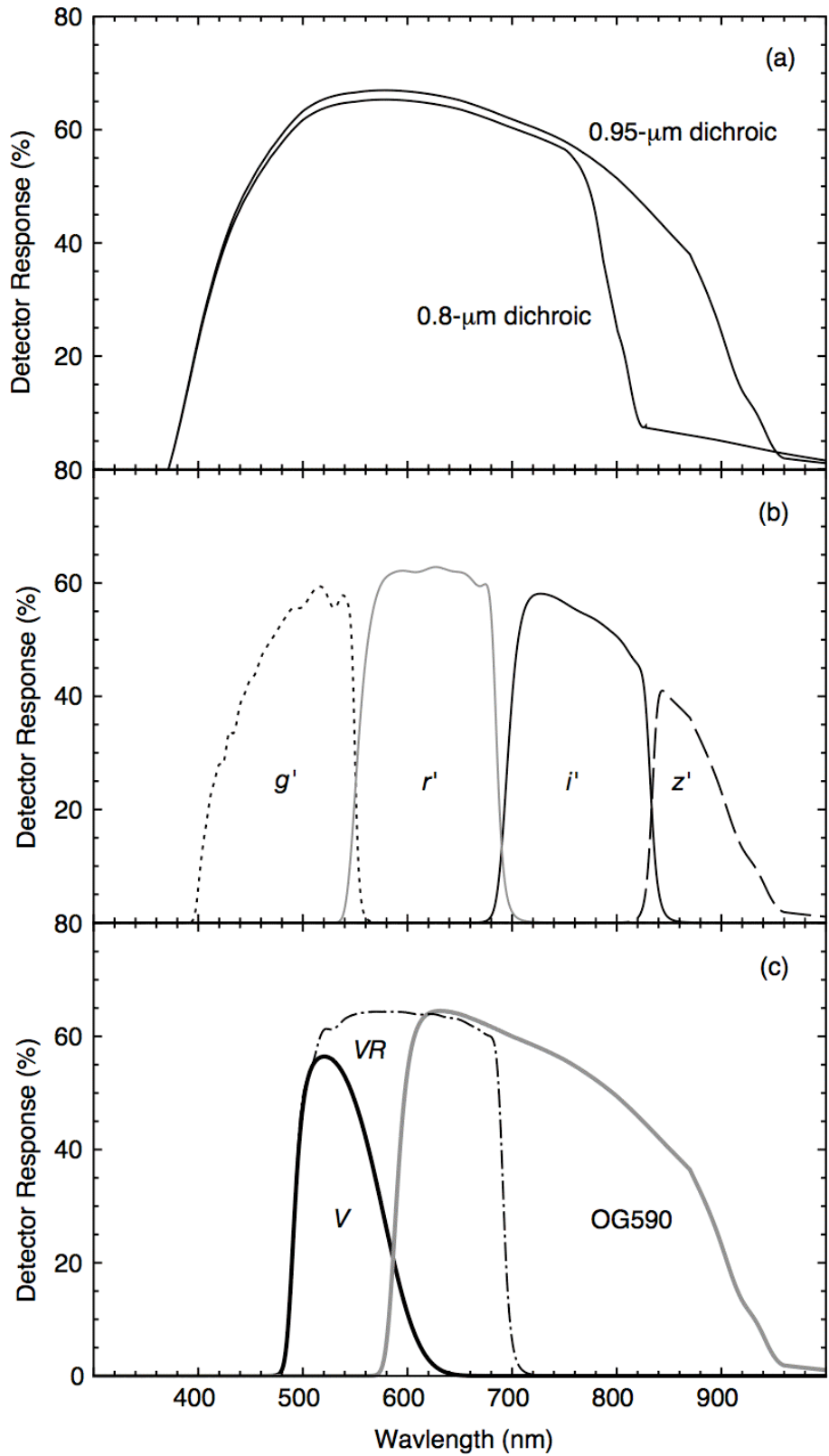

Figure 4. Gulbis *et al*.

  

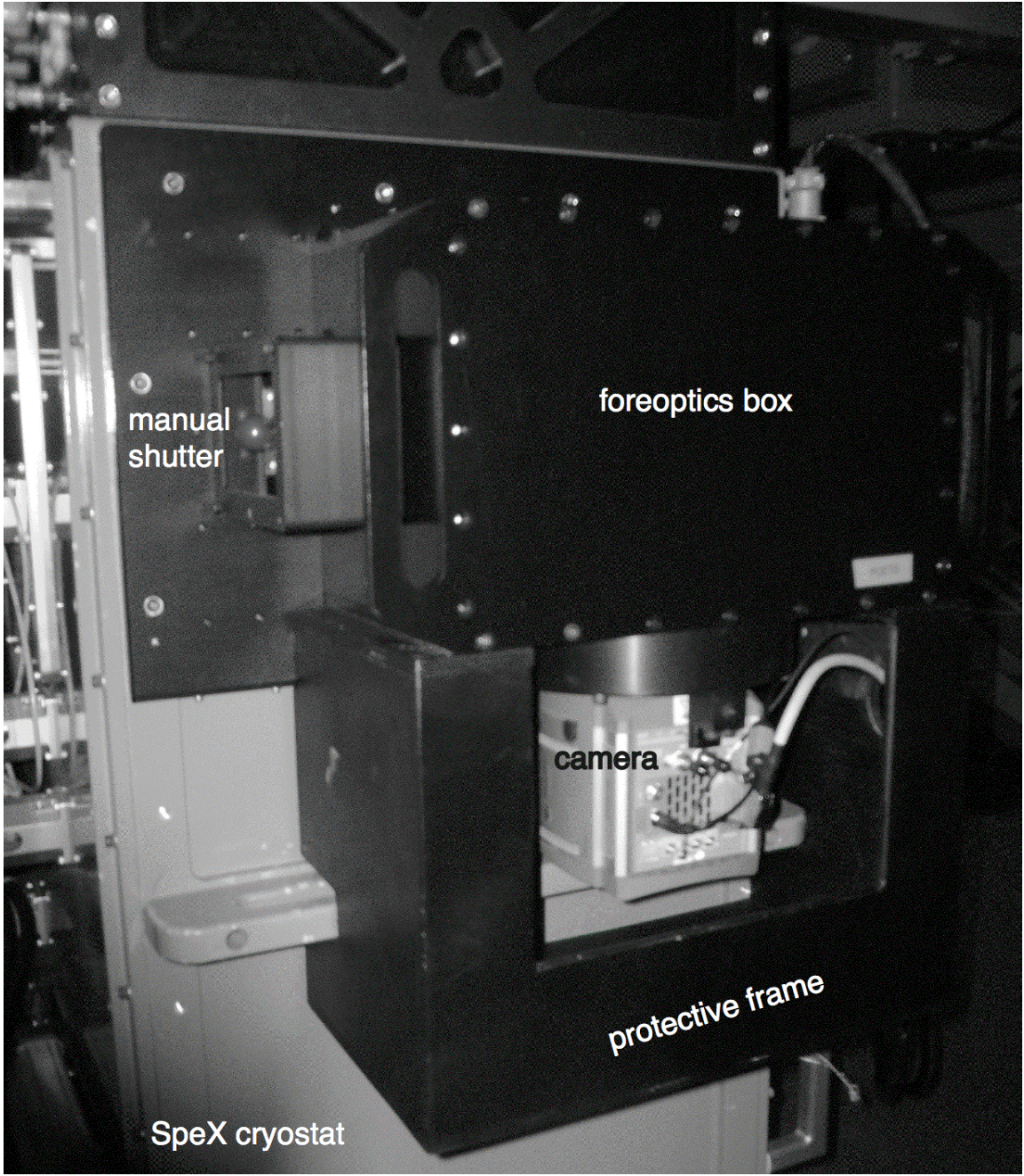

Figure 5. Gulbis *et al*.

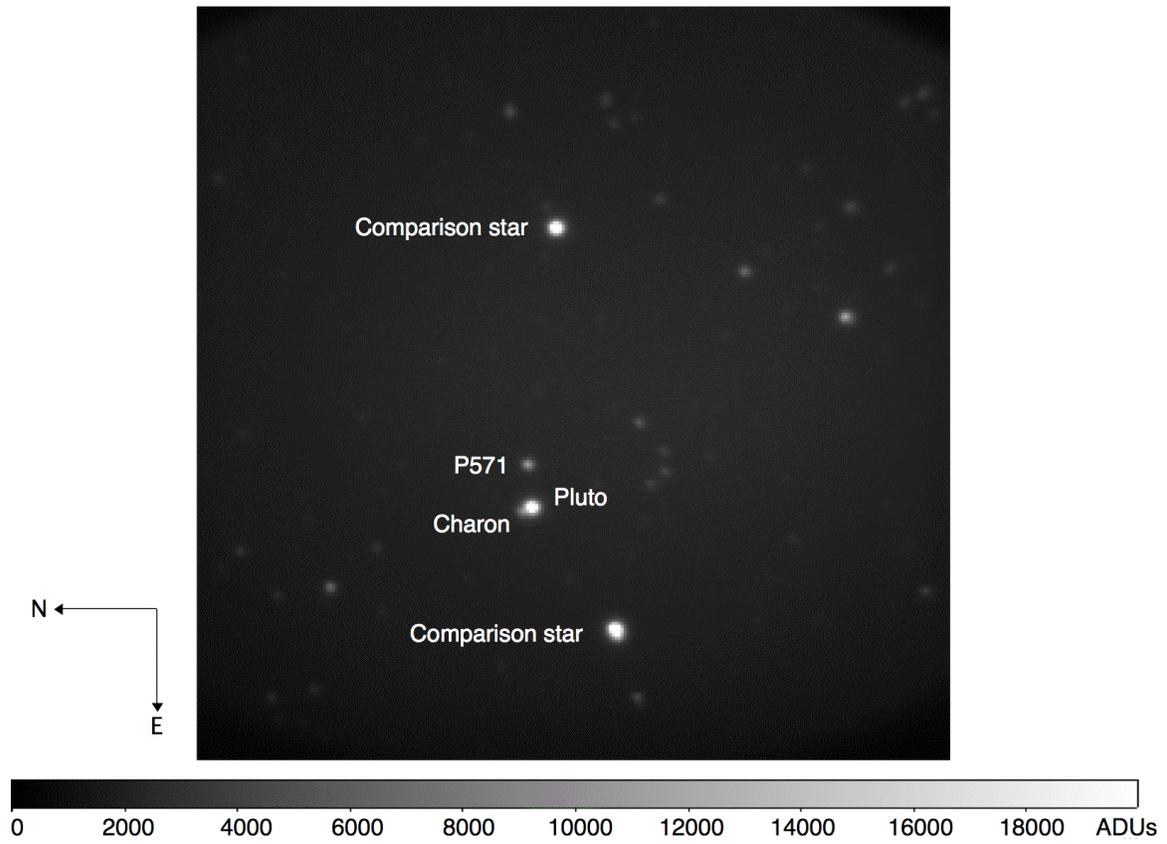

Figure 6. Gulbis *et al*.

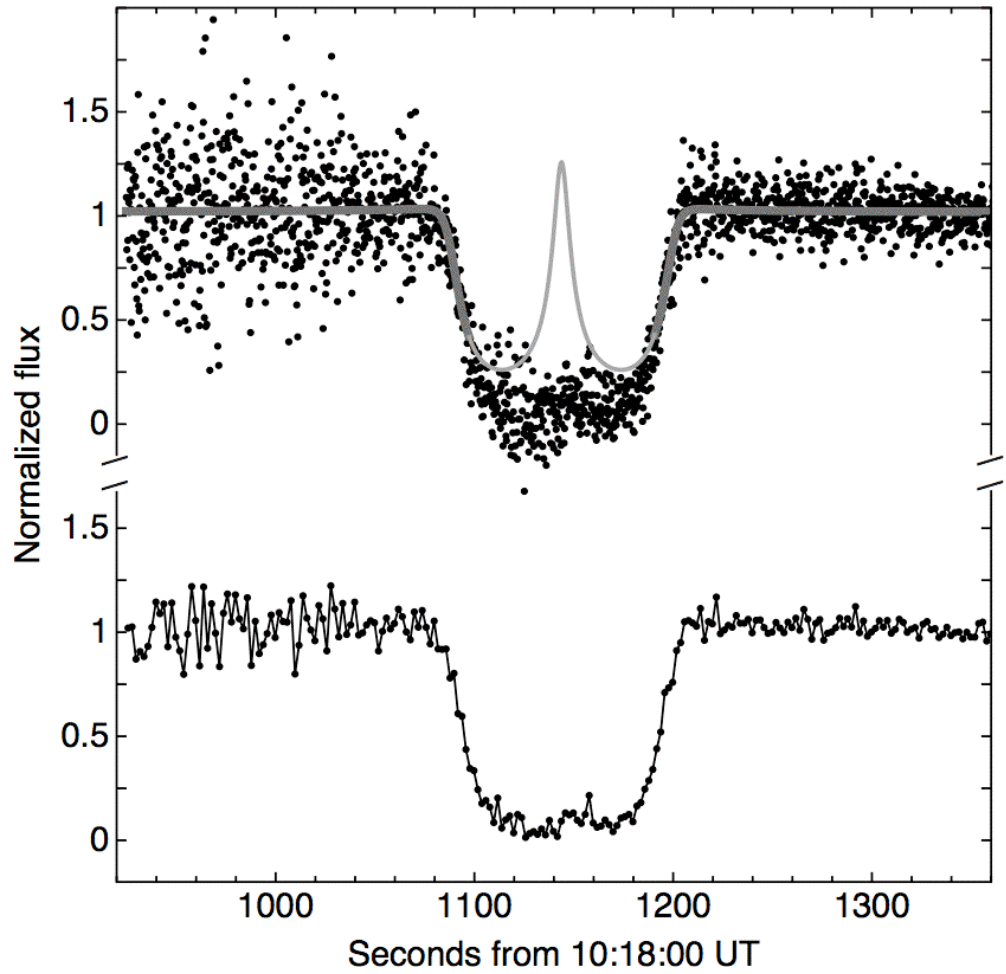

Figure 7. Gulbis *et al*.

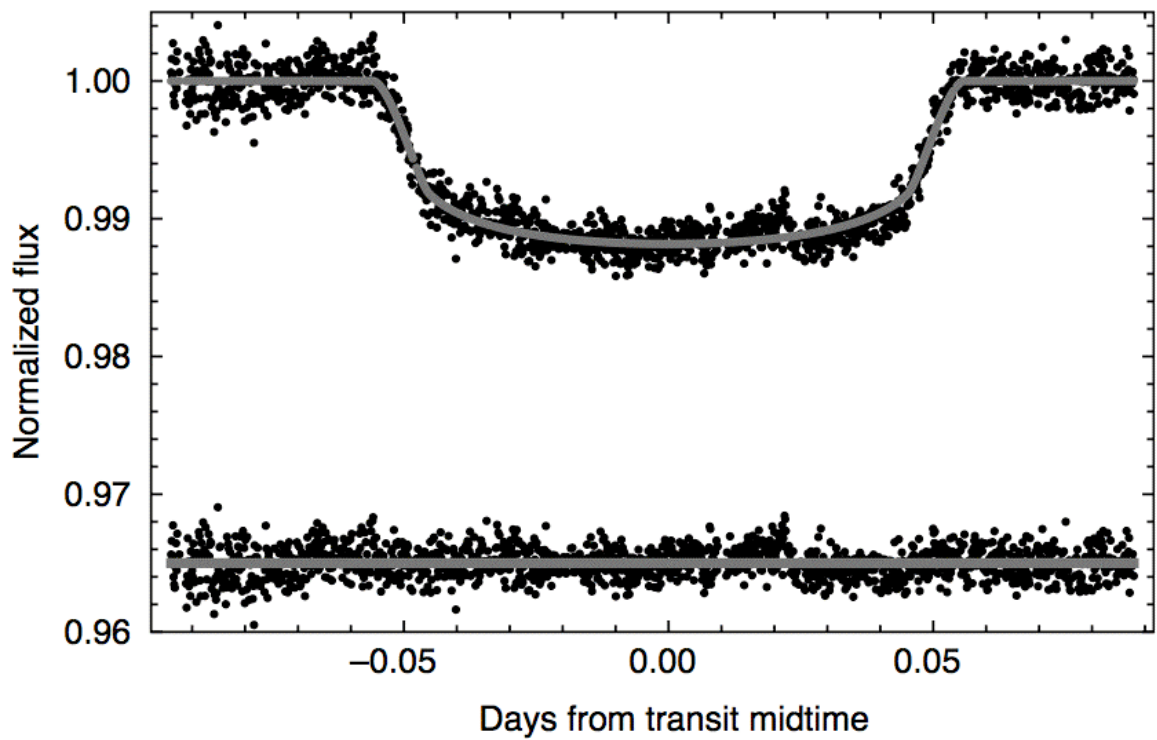

Figure 8. Gulbis *et al*.